\documentclass[twocolumn,prl,amsmath,amssymb,amsfonts,groupedaddress,floatfix,showpacs]{revtex4-1}

\usepackage[dvips]{graphics}
\usepackage{subfigure}
\usepackage{graphicx}
\usepackage{amsmath}
\usepackage{color}

\usepackage{graphicx}

\begin{document}


\title{Coexistance of giant tunneling electroresistance and magnetoresistance in
an all-oxide magnetic tunnel junction}

\author{Nuala Mai Caffrey}
\author{Thomas Archer}
\author{Ivan Runnger}
\author{Stefano Sanvito}
\affiliation{School of Physics and CRANN, Trinity College, Dublin 2, Ireland}

\date{\today}

\begin{abstract}
We demonstrate with first-principles electron transport calculations that large tunneling magnetoresistance (TMR) and tunneling electroresistance 
(TER) effects can coexist in an all-oxide device. The TMR originates from the symmetry-driven spin filtering provided by the insulating BaTiO$_3$ 
barrier to the electrons injected from SrRuO$_3$. In contrast the TER is possible only when a thin SrTiO$_3$ layer is intercalated at one of the
SrRuO$_3$/BaTiO$_3$ interfaces. As the complex band-structure of SrTiO$_3$ has the same symmetry than that of BaTiO$_3$, the inclusion of
such an intercalated layer does not negatively alter the TMR and in fact increases it. Crucially, the magnitude of the TER also scales with the thickness of 
the SrTiO$_3$ layer. The SrTiO$_3$ thickness becomes then a single control parameter for both the TMR and the TER effect. This protocol offers a 
practical way to the fabrication of four-state memory cells.
\end{abstract}


\maketitle

\label{Introduction}

Epitaxial magnetic tunnel junctions (MTJs), displaying giant tunneling magnetoresistance (TMR) at room temperature \cite{TMR1,TMR2}, represent 
the enabling technology for ultra-high density magnetic data storage. In MTJs the insulating barrier plays a dual role; it magnetically decouples the 
two ferromagnetic (FM) electrodes so that their magnetizations, $M$, can be arranged either parallel or antiparallel to each other, but it can also act 
as a spin filter, if epitaxially grown. This is due to the wavefunction symmetry selective decay of tunneling electrons across a crystalline insulator. As 
the two spin-manifolds of the Fermi surface of a FM metal present different symmetries, such symmetry selectivity translates into spin selectivity. This 
is the case for Fe/MgO(001), where the barrier is more transparent to the tunneling of electrons with $\Delta_1$ symmetry. These are present in Fe only 
for the majority spin \cite{Butler,Mathon}, so that the Fe/MgO(001) stack effectively behaves as a half-metal. 

In conventional MTJs, however, the insulating barrier is a passive element, i.e., its electronic structure cannot be changed by external  stimuli. A different 
situation is encountered when using a ferroelectric (FE) material. A FE is intrinsically insulating and at the same time possesses a macroscopic order, the 
electrical polarization, $P$. When FE materials are incorporated into a tunnel junction, one expects the junction resistance to become dependent on the 
direction of $P$ with respect to the layer stack, an effect known as tunneling electroresistance (TER) \cite{TER}. It then becomes natural to think about 
devices combining materials with both FM and FE orderings \cite{ScottNew}. Here, one can exploit the possibility of manipulating the two independent order 
parameters, $P$ and $M$, by means of their conjugate fields, namely the electric and magnetic fields. The fabrication of FE random access memories 
with non-destructive reading \cite{GajekNatMat6} demonstrates the potential of such an approach.

Although it is possible, at least in principle, to obtain a large TMR in MTJs with FE barriers \cite{Nuala,TsymbalTEMMR}, it is sensibly more complicated 
to obtain a large TER. The key ingredient for a MTJ to show TER is that it should exhibit inversion symmetry breaking. This is almost always the case in 
real devices as unintentional disorder breaks the symmetry. However, disorder is hardly controllable. Furthermore, even when the entire junction termination 
is different at either side of the insulating barrier the TER appears relatively modest \cite{TsymbalTEMMR}. A second strategy uses two magnetic electrodes 
made of different metals and thus different abilities to screen surface charges \cite{TER}. In typical metals with high carrier mobility, however, the screening 
length is short and the surface charge is strongly localized at the interface. The resulting potential profile thus remains approximately mirror-symmetric upon 
polarization reversal and the expected TER is small. Even for Fe/BaTiO$_3$/La$_{0.67}$Sr$_{0.33}$MnO$_3$ junctions, where
La$_{0.67}$Sr$_{0.33}$MnO$_3$ is quite a poor metal, a TER of only 37\% has been observed~\cite{Garcia26022010}.

One can bring the concept of having different screening lengths at the two sides of the FE layer to the extreme by including a second insulator (INS) 
in the stack, i.e., by considering a FM/INS/FE/FM junction. An interesting example of such structure is when the INS is actually vacuum, as in a
scanning tunnel microscopy experiment. Here, an extremely large TER has already been measured \cite{crassous:042901,Maksymovych12062009}. 
In this letter we demonstrate, by first principles electronic transport calculations, that the same effect can be achieved in an all-oxide solid state
device. In particular, we show that intercalating a few monolayers of SrTiO$_3$ into a SrRuO$_3$/BaTiO$_3$/SrRuO$_3$ junction creates an additional 
potential barrier that is switchable with the FE polarization of BaTiO$_3$. This not only amplifies the TER, but also makes it exponentially dependent 
on the SrTiO$_3$ thickness. Furthermore, as SrTiO$_3$ is electronically very similar to BaTiO$_3$ and thereby provides comparable spin-filtering for 
SrRuO$_3$ \cite{Nuala,TsymbalTEMMR}, the junction also displays a remarkably large TMR. 

\label{Methology & Structure}
The electronic structure of the junction is calculated by using density functional theory (DFT) as numerically implemented in 
the \textsc{siesta} code \cite{Soler_et_al:2002}. Structural relaxation is performed with the Perdew-Burke-Ernzerhof generalized 
gradient approximation (GGA) \cite{PhysRevLett.77.3865} to the exchange and correlation functional, with a 6$\times$6$\times$1 
$k$-point Monkhorst-Pack mesh and a grid spacing equivalent to a plane-wave cutoff of 800~eV. In contrast, for the electronic 
properties and the transport we use the atomic self-interaction correction (ASIC) scheme \cite{Pemmaraju/Sanvito:2007} built over 
the local spin density approximation. ASIC has been previously found to improve the electronic properties of bulk BaTiO$_3$~\cite{Tom} 
and SrRuO$_3$~\cite{rondinelli:155107} and it is vital in transport calculations where one has to ensure a good band alignment between 
dissimilar materials~\cite{PhysRevB.83.235112}. Unfortunately, even in its variational form~\cite{VPSIC} the ASIC scheme does not 
describe the FE phase accurately enough, so that a compromise is required; the GGA is used for the relaxation and the ASIC for the transport 
calculations. Electron transport is computed with the \emph{ab initio} code~\textsc{smeagol} \cite{Smeagol1,Smeagol2,Smeagol3}, which combines 
DFT with the non-equilibrium Green's functions scheme. \textsc{smeagol} uses \textsc{siesta} as its DFT platform so that the same convergence parameters 
are employed for the transport, except for the $k$-point sampling where we consider a much larger 100$\times$100$\times$1 mesh. 

\label{Displacements and MAV}
The supercell considered here comprises 6 BaTiO$_3$ unit cells ($\sim$~2.5~nm) and 3 unit cells of SrRuO$_3$ attached at each side to 
function as electrodes. Furthermore, we intercalate a thin SrTiO$_3$ layer between BaTiO$_3$ and SrRuO$_3$ at one side of the junction so that that 
the final stack is (SrO-RuO$_2$)$_3$/(SrO-TiO$_2$)$_m$/(BaO-TiO$_2$)$_6$/(SrO-RuO$_2$)$_3$, where $m = 0, 1, 2$. The in-plane lattice parameter 
is set to that of bulk SrTiO$_3$ (3.95\AA) to mimic the effect of a SrTiO$_3$ substrate. This applies compressive strain to both SrRuO$_3$ and
BaTiO$_3$ and in doing so increases the polarization of BaTiO$_3$. The Berry phase method gives a GGA polarization of 43.8$\mu$C/cm$^2$ for 
the bulk ($c/a$ = 1.05) and 48.1$\mu$ C/cm$^2$ for the strained structure ($c/a$ = 1.08). Note that the GGA systematically overestimates the polarization 
of FE oxides, but such a detail does not affect our conclusions.

We consider two alternative geometries for the junction, characterized by the
BaTiO$_3$ polarization pointing in opposite directions. 
Both geometries are relaxed to a tolerance of 40~meV/\AA~(less than 
4~meV/\AA\ for the $m = 0$ case). When BaTiO$_3$ is included in the capacitor
structure the displacements at the center of the 
supercell correspond to a polarization of 35.5$\mu$C/cm$^2$, i.e., 
sensibly reduced from its bulk value. 
\begin{figure}[h!]
\includegraphics[trim=0cm 0cm 0cm 0cm, clip, width=0.5\textwidth]{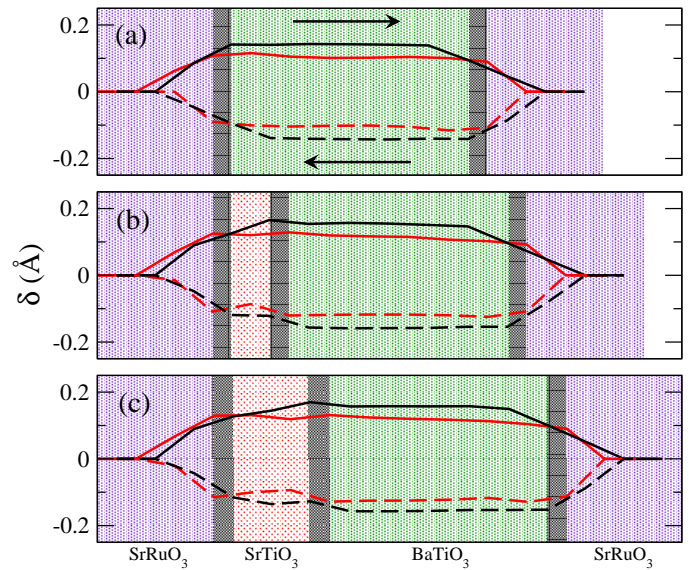}
\caption{\label{displacements}Relative cation-oxygen displacements along the $z$-axis (the junction stack direction) for the fully relaxed 
structure: (a) $m=0$, (b) $m=1$ and (c) $m=2$. The black (red) line corresponds to displacements in the BO$_2$ (AO) planes, of ABO$_3$. 
The solid (dashed) lines indicate $P_\rightarrow$ ($P_\leftarrow$).}
\end{figure}

An indication of the polarization structure is obtained from Fig.~\ref{displacements}, where we show the atomic displacements, $\delta$, along 
the MTJ stack. Here $\delta = (z_\mathrm{cation} - z_\mathrm{O})$, where $z_\mathrm{cation}$ and $z_\mathrm{O}$ denote respectively the 
cation and O position in a particular plane. As such, $\delta > 0$ defines a structure with the polarization pointing parallel to the substrate 
normal and away from the intercalated SrTiO$_3$ layer ($P_\rightarrow$), while $\delta < 0$ define 
a structure with the polarization pointing in the opposite direction ($P_\leftarrow$). Clearly, as far as the displacement is concerned there are no 
significant differences between the intercalated SrTiO$_3$ and BaTiO$_3$, which means that SrTiO$_3$, an incipient FE, takes on the FE distortion 
of BaTiO$_3$. This is valid only for thin SrTiO$_3$ films ($m=1, 2$) while we expect that thicker layers will lose the FE state.
\begin{figure}[h]
\includegraphics[trim=0cm 0cm 0cm 0cm, clip, width=0.5\textwidth]{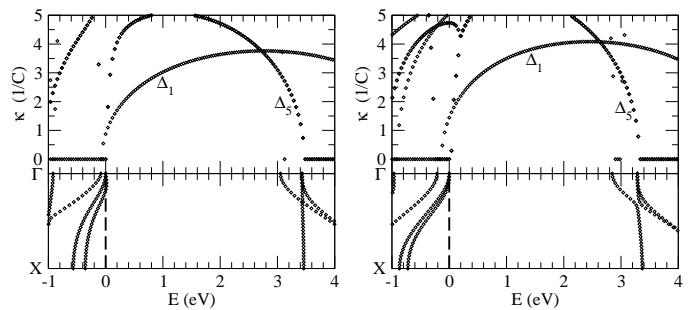}
\caption{\label{bto_sto_complex}Complex and real band structure of bulk SrTiO$_3$ (left panel) and bulk BaTiO$_3$ (right panel), calculated for the
FE structure constrained to the in-plane lattice parameter of SrTiO$_3$.}
\end{figure}

\label{TMR}
The electronic structures of SrTiO$_3$ and BaTiO$_3$ are also rather similar to each other as shown in Fig.~\ref{bto_sto_complex}, where
one can observe that the real band structures of the two materials almost coincide. Furthermore, and more importantly here, the symmetry of the 
complex part of the band structure is identical in both materials, with a $\Delta_1$ symmetry band dominating the lower energy part of the band gap 
and a $\Delta_5$ one defining the region near to the conduction band. We then expect that intercalating a SrTiO$_3$ layer will give a MTJ with the 
same spin-filtering properties of the SrRuO$_3$/BaTiO$_3$/SrRuO$_3$ stack \cite{Nuala}. Here, we use the ``optimistic'' TMR ratio,
$R_\mathrm{TMR}=\frac{G_{\uparrow\uparrow}-G_{\uparrow\downarrow}}{G_{\uparrow\downarrow}}$,
where $G_{\uparrow\uparrow}$ ($G_{\uparrow\downarrow}$) is the total conductance for the parallel (antiparallel) orientation of the magnetization. 
The TMR is found to increase with SrTiO$_3$ thickness due to the increasing length of the spin-filtering barrier. In particular, for $m=2$ 
at zero-bias $R_\mathrm{TMR}$ exceeds 10$^{8}$\% (the actual value depending on the polarization direction), meaning that at these thicknesses the 
barrier acts as an almost perfect spin-filter. 

We now discuss the TER effect in the junction by first looking at the electrostatic potential profile. In order to sustain the internal electric
field associated with a FE material, the electrostatic potential profile must display a finite slope. Concurrently, assuming that no bias is applied
across the junction and that the two electrodes are made of identical materials, i.e., they have the same Fermi energy, the average
potential in the electrodes should be identical. As a consequence, it is necessary that surface charge forms at the interface between the
FE layer and the metallic electrodes. This creates a depolarizing field so that the potential across the interface can be matched
and also sets the critical thickness for the onset of the FE state in a thin film \cite{nature01501}.
\begin{figure}[ht!]
\includegraphics[width=0.45\textwidth]{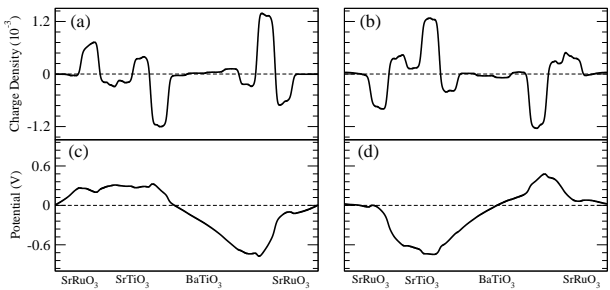}
\caption{\label{charge}Difference in charge distribution and electrostatic potential profile through the $m=2$ junction. These are the differences between 
the relative quantities (charge density and electrostatic potential) as calculated for the FE and centrosymmetric structures and averaged 
over the plane perpendicular to the junction stack. The left (right) panel is for $P_\rightarrow$ ($P_\leftarrow$).}
\end{figure}

In Fig.~\ref{charge} we present both the charge density and the electrostatic potential profile across the $m=2$ junction. These are obtained as the planar 
average of the difference between the relevant quantity calculated for the centrosymmetric and FE configurations. The atomic oscillations thereby cancel
and one is left with the modifications of the charge density and the potential due to the onset of the FE phase. In general we observe that charge density 
of opposite sign forms at either side of the FE layer resulting in the expected potential difference. As we move into the metallic layers at the 
BaTiO$_3$/SrRuO$_3$ interface an additional peak in the charge density can be seen, which acts as a depolarization charge and brings the potential back 
to zero. In contrast, at the SrTiO$_3$/BaTiO$_3$ interface there are not sufficient screening charges so that the depolarization charge forms instead at the 
metallic SrRuO$_3$ electrode. This leaves the potential in SrTiO$_3$ pinned to that at the interface with BaTiO$_3$. Thus, when one reverses $P$ the 
potential in the SrTiO$_3$ layer is rigidly shifted. Note that, as a consequence of such charge distribution, the average electrostatic potential 
in SrTiO$_3$ remains flat despite the ionic displacement.
\begin{figure}[ht!]
\includegraphics[trim=1cm 1cm 2cm 2cm,clip , width=0.40\textwidth]{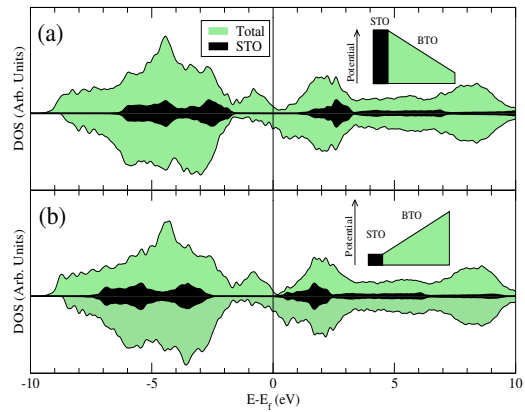}
\caption{\label{DOS}Total density of states (DOS) for the $m=2$ junction (green shaded area) and DOS projected over the SrTiO$_3$ layer (solid black area) 
for the two different polarization directions, namely (a) $P_\rightarrow$ and (b) $P_\leftarrow$. Note the rigid shift in the SrTiO$_3$ DOS as $P$ is reversed. 
In the insets we show a schematic of the barrier profile for both polarization directions. }
\end{figure}

This rigid shift in the potential can be appreciated by looking at Fig.~\ref{DOS}, where we show that the density of states (DOS) projected onto the SrTiO$_3$ 
layer is rigidly displaced by the reversal of the BaTiO$_3$ polarization direction. In particular, for the $P_\leftarrow$ configuration the SrTiO$_3$ conduction band 
edge is considerably closer to the junction Fermi level, $E_\mathrm{F}$, than for the $P_\rightarrow$ case. This means that the height of the SrTiO$_3$ potential 
barrier presented to the tunneling electrons changes according to the direction of $P$. In summary, the overall scattering potential appears as follows: for
$P_\rightarrow$ there is a high barrier in SrTiO$_3$ followed by a triangular barrier in BaTiO$_3$, which decreases as one moves away from the TiO$_2$/BaO 
interface [see insets of Fig.~\ref{DOS}(a)]. In contrast, for $P_\leftarrow$ the SrTiO$_3$ barrier is small while the triangular BaTiO$_3$ barrier increases 
away from the interface. As a consequence, tunneling through BaTiO$_3$ is essentially insensitive to the polarization direction, as the two triangular 
barriers are identical. However, the barrier height across SrTiO$_3$ changes significantly with the $P$ direction. Such a polarization-dependent change
in the SrTiO$_3$ barrier is the cause of the TER effect in this junction.

In Table~\ref{tab:G_TER} we summarize our transport results. In particular, we present the junction conductance at zero bias for the two different $P$
directions (either $\leftarrow$ or $\rightarrow$) and the two different magnetic arrangements of the electrodes (either parallel $\uparrow\uparrow$ or
antiparallel $\uparrow\downarrow$ orientation) for $m=0, 1$ and $2$. In the table, in addition to $R_\mathrm{TMR}$, we also report the figure of merit for 
the TER effect, namely the TER ratio $R_\mathrm{TER}^{\sigma\sigma^\prime}=\frac{G_\leftarrow^{\sigma\sigma^\prime}
-G_\rightarrow^{\sigma\sigma^\prime}}{G_\rightarrow^{\sigma\sigma^\prime}}$.
Note that the TMR is now dependent on the polarization direction so the ratio is defined as
$R_\mathrm{TMR}^\alpha=\frac{G^{\uparrow\uparrow}_\alpha-G^{\uparrow\downarrow}
_\alpha}{G^{\uparrow\downarrow}_\alpha}$, being $G_\alpha^{\sigma\sigma^\prime}$ the junction conductance for the $\sigma\sigma^\prime$ magnetic 
configuration and $P$ pointing in the $\alpha$ direction.

\label{T and TER}
\begin{table}
\begin{ruledtabular}
\begin{tabular}{ccccccc}
$m$ & ${\sigma\sigma^\prime}$		&
$G^{\sigma\sigma^\prime}_\rightarrow$ 	&
$G^{\sigma\sigma^\prime}_\leftarrow$ 	& 
$R^{\sigma\sigma^\prime}_\mathrm{TER}$ & $R_\mathrm{TMR}^\rightarrow$	&
$R_\mathrm{TMR}^\leftarrow$ \\ \hline
0&	$\uparrow\uparrow$ &	4.05x10$^{6}$	&	4.06x10$^{6}$	 &
0.31	&	&  \\
\hline
1&	$\uparrow\uparrow$ &	6.82x10$^{4}$	&	9.49x10$^{4}$	 &
39.07		& &  \\
\hline
2&	$\uparrow\uparrow$ &	2.86x10$^{3}$	&	8.95x10$^{3}$	 &
212.84		& & \\
2&	$\uparrow\downarrow$ &	1.12x10$^{-4}$	& 1.83x10$^{-3}$ & 1533.93 &
2.5x10$^9$ & 4.75x10$^8$ \\ 
\end{tabular}
\end{ruledtabular}
\caption{\label{tab:G_TER} Layer conductance (in units of $\Omega^{-1}$cm$^{-2}$) and both TMR and TER ratios (in \%) for different
$m$. Here $G_\alpha^{\sigma\sigma^\prime}$ is the layer conductance for the magnetic configuration $\sigma\sigma^\prime$ and the electrical 
polarization pointing in the $\alpha$ direction. Note that the TER depends on the magnetic configuration of the junction and the TMR 
depends on the electrical configuration.
}
\end{table}

Firstly, it can be observed that our MTJ sustains a very robust TMR regardless of the direction of the polarization vector. This is simply a consequence 
of the spin-filtering effect and of the fact that the electronic structure of SrTiO$_3$ and BaTiO$_3$ are rather similar. More interesting is the dependence
of the TER on the SrTiO$_3$ barrier. Since in our junction the TER originates from a change in the SrTiO$_3$ barrier height, the effect is expected to be 
magnified by increasing the barrier width, i.e., the SrTiO$_3$ layer thickness. This is indeed the case, as demonstrated by the dependence of
$R_\mathrm{TER}$ on $m$ reported in Table~\ref{tab:G_TER}. In particular, we find  that $R_\mathrm{TER}\sim0$ for $m=0$, i.e., when there is no intercalated
SrTiO$_3$. It then increases drastically for $m=1$ and $m=2$. This increase is, in fact, exponential in $m$, and it goes as $e^{-(\Delta^\leftarrow-\Delta^\rightarrow)m}$,
where $\Delta^\alpha$ is the SrTiO$_3$ barrier height in the $P_\alpha$ configuration. This is an important result, as it demonstrates
that the TER can be tuned to a great degree by simply controlling the SrTiO$_3$ layer thickness.

It is also important to note that for a given ($m\ne0$) junction there are four very distinct conductance states depending on both the magnetization direction
of the electrodes and the polarization direction of the ferroelectric layer. This means that our proposed device can operate as a four state memory cell 
with four well-separated conductive states. Finally, one can quantify the dependence of the TMR on the $P$ direction by calculating the tunneling
electro-magneto resistance (TEMR) ratio, defined as $R_\mathrm{TEMR}=
\frac{R_\mathrm{TMR}^\rightarrow -
R_\mathrm{TMR}^\leftarrow}{R_\mathrm{TMR}^\leftarrow}$.
For $m=2$ we find $R_\mathrm{TEMR}=460$\%, a value which is comparable to those reported experimentally for 
Fe/BaTiO$_3$/La$_{0.67}$Sr$_{0.33}$MnO$_3$ thin film structures (ranging between
140\% and 450\%) \cite{Garcia26022010}.

\label{Conclusions}

In conclusion, we have discussed the effects of including a wide-band gap insulator into a MTJ based on a FE barrier.
We have demonstrated that in such a junction the tunneling barrier profile can be tuned by reversing the direction of the 
macroscopic electrical polarization. This results in a tunable TER effect which may coexist with a TMR effect. In particular, the 
choice of SrTiO$_3$ and BaTiO$_3$, which both offer excellent spin-filtering to spins injected from SrRuO$_3$, results 
in a device which also displays remarkably large TMR ratios. Importantly, both the TMR and the TER are tunable and 
increase with the SrTiO$_3$ layer thickness. As such our proposed stack offers a robust protocol for constructing devices
displaying simultaneous TER and TMR effects. These can operate as a four state memory element for data storage applications. 

This work is sponsored by Science Foundation of Ireland (07/IN.1/I945) and by the EU-FP7 (ATHENA and iFOX projects). 
IR is sponsored by the King Abdullah University of Science and Technology ({\sc acrab} project). Computational resources have 
been provided by the HEA IITAC project managed by TCHPC.

\end{document}